\documentstyle[12pt]{article}
\newcommand{\be}{\begin{equation}}
\newcommand{\ee}{\end{equation}}
\newcommand{\ii}{{\it i}}
\newcommand{\f}{\varphi}
\newcommand{\iii}{\int_{0}^{\infty}}
\begin{document}

\begin{titlepage}
\begin{flushright}
Z\"urich University Preprint

ZU-TH 32/93\ \ \ \ \
\end{flushright}

\vspace{20 mm}

\begin{center}
\huge

Einstein-Yang-Mills sphalerons and fermion number non-conservation

\vspace{5 mm}
\end{center}
\begin{center}
{\bf  Mikhail S. Volkov}\footnote{On  leave  from  Physical-Technical
Institute of the  Academy  of  Sciences  of  Russia,  Kazan  420029,
Russia}

\vspace{5 mm}
Institut    f\"ur    Theoretische    Physik    der    Universit\"at
Z\"urich-Irchel, Winterthurerstrasse 190, CH-8057 Z\"urich,

Switzerland

\end{center}
\vspace{1 cm}
\begin{center}
{\large Abstract}
\end{center}
\vspace{1 cm}

It  is  argued  that  the  sphaleron  solutions  appearing  in   the
Einstein-Yang-Mills theory are important in the transition processes
at extremely high energies. Namely,  when  the  energy  exceeds  the
sphaleron  mass,  the  existence  of  these  solutions  ensures
constructive interference on the set of the  overbarrier  transition
histories interpolating between distinct topological sectors of  the
theory. This has to lead to an enhancement of the transition rate and
the related  non-conservation of the fermion number.

\end{titlepage}
\newpage

     The discovery of the vacuum   periodicity   in   Yang-Mills
(YM) theory \cite{1} gave rise to the  theoretical  description   of
baryon number  non-conservation \cite{2}  caused  by  the  anomalous
violation   of chirality \cite{3}. The standard approach to
understanding this phenomenon uses transitions between
topologically distinct YM vacua divided by a
potential barrier.  In an unbroken YM  theory
 (like QCD) this barrier
has no definite height and  can  be  made  infinitely
small due to the scale invariance. In a spontaneously broken  theory
(like the  electroweak  one)  the  scale  invariance  is  explicitly
violated by the Higgs field vacuum expectation value,  and  the
potential
barrier acquires definite non-zero minimal height. The certain value
of this height is defined by the mass of the  electroweak  sphaleron
solution ``sitting'' on the top of the barrier  \cite{5}.  When  the
energy  is  small,  all  histories  interpolating  between  distinct
topological  sectors  encounter  the  barrier,  which  leads  to  the
tunneling suppression. On  the  other  hand,  when  the  energy  (or
temperature) exceeds the barrier height, the system may pass  freely
over the barrier, causing  an  efficient non-conservation  of  the
fermion number \cite{6,7}.

When the energy is very high, it is natural to expect  the symmetry
to be restored. However, another universal scale arises  naturally  at
the extremely high energies, namely, the gravitational Planck length.
It is interesting to consider the possibility of  the
scale  invariance being violated by
this  length.  As  is  known,   in   the
Einstein-Yang-Mills (EYM) theory there exist  regular  particle-like
solutions \cite{9}. Similarly to the  electroweak  sphaleron,  these
solutions are ``sitting''  on  the  top  of  the  potential  barrier
dividing distinct vacua, which  allows them also to be interpreted as
sphalerons \cite{10}. One may think that the masses of these  EYM
sphalerons, as in  the  electroweak  theory,
define the typical  value  of  energies  when  the  fermion  number
non-conservation becomes significant.

There  is,  however,  an  important  difference  between   the   EYM
sphalerons and their electroweak counterpart. It turns out that  the
masses of the EYM sphalerons do not define the {\it  minimal}  value
of the barrier height,  as  the  latter  may  be  arbitrarily  small
\cite{10}. This situation resembles  that  appearing  in  the  scale
invariant YM theory because gravity, while  violating  the
scale invariance and making the existence of the sphalerons possible,
may only reduce the potential energy. In other words, the profile of
the potential barrier is different in this case; there are  gaps
in the barrier,
and the sphalerons correspond to critical  saddle  points  on
the barrier surface  with  more  than  one  negative  direction  of
instability. This means that the overbarrier  passages  in  the  EYM
theory are always possible, not only when  the  energy  exceeds  the
sphaleron mass. At first  glance,  this  makes  the  physical
meaning of the EYM sphalerons and their role in the  enhancement  of
the transition rate at high energies unclear.

The purpose of this paper is to show that  the  EYM  sphalerons  are
nevertheless important  in  the  transition  processes  due  to  the
interference effects. Namely, we argue that when the energy  exceeds
the sphaleron mass, the existence of  the  critical  points  on  the
barrier surface ensures constructive interference on the set  of the
overbarrier transition histories. This has to lead to an  enhancement
of the  transition intensity.  When  the  energy  is  small,  the
overbarrier histories also exist,
however, the constructive interference  is  absent.
We conclude therefore that the rate of the fermion number  violating
processes increases when the energy exceeds the sphaleron mass.

Our aim is to estimate (in the lowest order WKB  approximation)  the
amplitude of the fermion number  violating  transition  in  the  EYM
theory at non-zero energy. The theoretical  tool  relevant  for  our
purposes has been developed by Bitar and Chang in  their  real  time
analysis of tunneling  transitions  in  the  flat  space  YM  theory
\cite{11}.  The  idea  of  this  method  is, first, to  define    a
one-parameter  continuous  family  of  static  field  configurations
in the configuration function space  of  a  theory,  such  that  this
family connects two vacuum sectors with different  winding  numbers.
Let us call such a family a vacuum-to-vacuum (VTV)  path.  Then  one
assumes that the transition between distinct vacuum sectors  of  the
theory proceeds along this single VTV path. This reduces the problem
to a one-dimensional  quantum-mechanical  task,  and  allows  one  to
find the corresponding WKB amplitude.  The  last
step is to take the sum over all VTV paths. Of course, in  order  to
be able to follow this program, we apply some approximations.

Consider the action of the  EYM  theory  with  the  $SU(2)$ gauge
group
\be
S_{EYM} = S_{G}+S_{YM},                                   \label{1}
\ee
with the gravitational part
\be
S_{G}=-{1\over 16\pi G}\int R\sqrt{-g}d^{4}x -
{1\over 16\pi  G}\oint_{\Sigma}
(g^{\mu\alpha}\Gamma^{\beta}_{\alpha\beta}-
g^{\alpha\beta}\Gamma^{\mu}_{\alpha\beta})d\Sigma_{\mu}, \label{2}
\ee
and the Yang-Mills contribution
\be
S_{YM}=-{1\over 2g^{2}}\int trF_{\mu\nu}F^{\mu\nu}\sqrt{-g}d^{4}x.
                                                        \label{3}
\ee
Here $G$ is Newton's constant, $g$ is the gauge  coupling  constant,
$F_{\mu\nu}=\partial_{\mu}A_{\nu}-\partial_{\nu}A_{\mu}-        {\it
i}[A_{\mu},A_{\nu}]$ is the  matrix  valued  gauge   field   tensor,
$A_{\mu}=A_{\mu}^{a}\tau^{a}/2$, and $\tau^{a}\  (a=1,2,3)$  are the
Pauli matrices. We use the  gravitational  action,  which  does  not
contain second derivatives of the  metric  and  reduces
to the well-known two-gamma action \cite{ll} when an  event  horizon
is absent; the surface term in Eq.(\ref{2})  is  non-covariant,  and
quasi-Cartesian coordinates are used for calculation of it  (we  are
working in an asymptotically flat spacetime) \cite{main}.
 Our  sign  conventions
used  are those of Landau \& Lifshitz \cite{ll}.

The  action  (\ref{1})  has  many   stationary   points,   including
solutions  with non-trivial space-time topology such as black  holes
\cite{12}. Consider a sector with spacetime manifolds  carrying  the
trivial $R^{4}$  topology.  Define  vacua  in  this  sector  as  the
stationary points of the action with zero ADM  energy.   As  follows
from the positive energy theorem \cite{16}, such vacua must  have  a
flat metric and hence the YM field is a pure gauge. In the $A_{0}=0$
gauge, one can represent the vacuum fields as
\be
A_{j}(\vec{x})={\it  i}U\partial_{j}U^{-1}, \ \ \
g_{\mu\nu}(\vec{x})=\eta_{\mu\nu},                     \label{6}
\ee
where $\eta_{\mu\nu}$ is the flat spacetime  metric,  $U(\vec{x})$  is a
$SU(2)$ valued function, and  spatial  index  $j$  runs  over  1,2,3
$(\vec{x}  \equiv  x^{j})$.  As  is   known,   if    the    function
$U(\vec{x})$  in  Eq.(\ref{6})  satisfies  the following  additional
condition \cite{1}:
\be
\lim_{|\vec{x}|\rightarrow\infty}U(\vec{x}) = 1,          \label{11}
\ee
then all such vacua may be classified in terms  of  integer  winding
numbers of the gauge field:
\be
k = \int K^{0}d^{3}x  =  {1\over  24\pi^{2}}  \int
\varepsilon^{ijk}trU\partial_{i}U^{-1}U\partial_{j}U^{-1}U\partial_{k}
U^{-1} d^{3}x.                                            \label{10}
\ee
(One  should  note  that  restricting  ourselves  only  to  spacetime
manifolds with the $R^{4}$ topology, we exclude  from  consideration
all  non-trivial  topological  effects  of   gravity.   A   complete
investigation  of these effects  requires inclusion of  vacua   with
non-zero  topological  indices,  both  for  the  gauge  and   the
gravitational fields.)

The transitions between the sectors with different  winding  numbers
are of interest because they are accompanied  by  the  corresponding
change of the number of massless fermions due to the axial  anomaly.
To estimate the amplitude of  such  a  transition,  we  introduce  a
family of the VTV paths in  the  EYM  configuration  function  space
\cite{10}:
\be
 \{A_{j}(\vec{x},     \lambda);    \left.        g_{\mu\nu}(\vec{x},
\lambda)\}
\right|_{K(r)}. \ \ \ \                                   \label{35}
\ee
This family is labeled by a function $K(r)$ satisfying the following
conditions
\be
K(0)=1,\ \ K(\infty)=-1;                                  \label{23}
\ee
each path of the family is parameterized by $\lambda\in [0,\pi]$. The
explicit form of the gauge field related to a path is chosen to be
\be
A_{\mu}(t,\vec{x})=\frac{\ii}{2}(1-K(r))U\partial_{\mu}U^{-1},\ \ \
U=exp(\ii\lambda n^{a}\tau^{a}),                         \label{22}
\ee
where   $n^{a}=(sin\vartheta cos\varphi, sin\vartheta   sin\varphi,
cos\vartheta)$ is the unit vector. The gravitational field is
\be
ds^{2} = R^{2}_{g} \{(1-\frac{2m}{r})\sigma^{2}dt^{2} -
\frac{dr^{2}}{1-2m/r} - r^{2}(d\vartheta^{2} + sin^{2}\vartheta
d\varphi^{2})\},                                      \label{13}
\ee
with metric function $m(r)$ and $\sigma(r)$ being
$$
\sigma  (r)   =   exp\{   -2sin^{2}\lambda\int_{r}^{\infty}   K'^{2}
\frac{dr}{r}\}, $$
\be
m(r)=\frac{sin^{2}\lambda}{\sigma(r)}\int_{0}^{r}     (K'^{2}      +
sin^{2}\lambda\frac{(K^{2}-1)^{2}}{2r^{2}})\sigma dr,    \label{33}
\ee
where $R_{g}=\sqrt{4\pi}l_{pl}/g$ is the only  dimensional  quantity
in the  problem  ($l_{pl}$  being  Planck's  length).  These  fields
satisfy the single non-trivial constraint equation appearing in  the
case:
\be
G^{0}_{0}=8\pi GT^{0}_{0}.                        \label{constr}
\ee
When $\lambda$ runs from  zero  to  $\pi$  the  fields  interpolate
between the vacuum values (\ref{6}). If  $K(r)$  is  a  sufficiently
smooth     function     then     the     geometry     defined     by
Eqs.(\ref{13}),(\ref{33}) is everywhere regular  and  asymptotically
flat. The corresponding ADM mass is
\be
M = \lim_{r\rightarrow\infty}m(r).                       \label{34}
\ee
Our  basic  approximation  is  the  assumption  that   the   quantum
transition between distinct vacuum sectors  occurs  only  along  the
paths (\ref{35})-(\ref{33}). Notice that Bitar and Chang's  approach
gives an adequate description of the transition process provided one
is able to take into account {\it all\ } VTV paths.   But
the   main advantage of the method, perhaps,    is   that   it   may
give  a  good description of the transition even if one  takes  only
some particular set of these paths \cite{11}.

To estimate the transition amplitude, let us  first  choose  a  path
from the family (\ref{35}). To find the partial amplitude related to
this path, allow for the parameter  $\lambda$  to  depend  on  time:
$\lambda\rightarrow\lambda(t)$, and  calculate  the  action  of  the
fields. It is worth noting at this stage that when the time dynamics
is
introduced, one may show explicitly that all paths (\ref{35}) indeed
interpolate between vacua  with  distinct  winding numbers. To see
this, suppose that
$\lambda(t_{0})=0$, $\lambda(t_{1})=\pi$, then the direct calculation
of the gauge
invariant Pontryagin index for the gauge field Eq.(\ref{22}) yields
\cite{10,main}
\be
\nu  =  {1\over   16\pi^{2}}   \int_{t_{0}}^{t_{1}}dt\int
trF_{\mu\nu}\tilde{F}^{\mu\nu}
\sqrt{-g}d^{3}x = 1 ,                                      \label{nu}
\ee
so the time evolution along such a path has to change the winding
number.
Also, when $\lambda$ in Eq.(\ref{22}) depends  on  time,
the non-zero time component of the gauge field  arises.
Passing to the $A_{0}=0$ gauge, one obtains explicitly
$$
A_{\mu}=\ii \frac{1-K}{2}U_{+}\partial_{\mu}U_{+}^{-1}+
\ii \frac{1+K}{2}U_{-}\partial_{\mu}U_{-}^{-1},
$$
\be
U_{\pm}=exp(\ii\lambda (K\pm 1) n^{a}\tau^{a}/2), \label{temporal}
\ee
which is zero when $\lambda=0$, and a pure gauge
with unit winding number when $\lambda=\pi$.

Let  us  pass  to  the  calculation   of   the   action.   Inserting
Eqs.(\ref{22})-(\ref{33}) with $\lambda=\lambda(t)$  into  (\ref{1})
and performing direct but somewhat lengthy calculations \cite{main},
on may represent the result in the form
\be
S_{EYM}=\frac{4\pi}{g^{2}}\int
(\frac{\mu(\lambda)}{2}\dot{\lambda}^{2}-U(\lambda))\ dt,\label{43}
\ee
where
\be
\mu(\lambda)\equiv\mu[K(r),\lambda]=
\iii\frac{r^{2}}{\sigma}(K'^{2}+2sin^{2}\lambda\
\frac{(K^{2}-1)^{2}}{r^{2}-2mr\ })dr,                    \label{40}
\ee
and
$$
U(\lambda)\equiv U[K(r),\lambda]=$$
\be
=sin^{2}\lambda\iii (K'^{2} +
sin^{2}\lambda\frac{(K^{2}-1)^{2}}{2r^{2}})
exp(-2sin^{2}\lambda\int_{r}^{\infty}K'^{2}\frac{dr}{r})dr.
                                                         \label{45}
\ee
In  these  expressions  quantities  $m,\  \sigma$  are   given    by
Eq.(\ref{33}),  and $K(r)$ being  a  function  specifying  the  path
under  consideration.  One  can  see  that   the   result   obtained
corresponds  to  the  action  of  an   effective    particle    with
position-dependent   mass,    $\mu(\lambda)$,    moving     in     a
one-dimensional external  potential  $U(\lambda)$.   This  potential
has the typical barrier shape: for each $K(r)$ it  vanishes for  the
vacuum  values,  $\lambda=0,\pi$,  and   reaches   a   maximum    in
between at $\lambda=\pi/2$ (the latter can be seen if we pass  to  a
new independent variable $z=r/sin\lambda$ under the  integration  in
Eq.(\ref{45})). Notice that the potential  coincides  with  the  ADM
mass  (\ref{34})  (that  is true  provided  that  the
geometry described by the metric
(\ref{13}) is everywhere regular \cite{main}).

Thus, we arrive at the one-dimensional barrier  transition  problem.
The corresponding one-dimensional Schr\"odinger equation then reads
\be
{\cal H}=\frac{p^{2}}{2\mu(\lambda)}+U(\lambda)=E,
                                                        \label{37}
\ee
with $p$ being the momentum conjugated to $\lambda$,  and  the
quantity $E$ has the sense of the energy of the asymptotically  free
quantum states (the operator ordering problem can be avoided in this
case \cite{11}). This allows us  to  write  down  the  corresponding
partial WKB transition amplitude as follows
\be
{\cal A}_{K(r)}\equiv Bexp\{ \ii\frac{4\pi}{g^{2}}\Phi[K(r)]\}=
Bexp\{ \ii\frac{4\pi}{g^{2}}
\int_{0}^{\pi}  d\lambda \sqrt{2\mu(\lambda)[E-U(\lambda)]}\},
                                                          \label{46}
\ee
where $B$ absorbs all other WKB factors, which  are  inessential for
our present considerations. The subscript $K(r)$ indicates that this
partial amplitude  relates   to   a   single   path   specified   by
$K(r)$. If the quantity $E-U(\lambda)$ is negative, then one  should
take the value of the square root lying in the  upper  half  of  the
complex plane.

The last step in estimating the transition amplitude is to represent
the  total  amplitude  as  the  sum  over  all  partial   amplitudes
(\ref{46}):
\be
{\cal A}=
\sum_{K(r)}{\cal A}_{K(r)}.                               \label{47}
\ee
This expression gives the amplitude of the winding  number  changing
transition in the EYM theory at arbitrary energy $E$. It  is  clear,
however, that taking the sum is virtually impossible. The  only  way
to estimate this  sum  is  to  make  use  of  the  stationary  phase
approximation. However, as was shown in Ref.\cite{11},  the  finding
of the exact stationary phase path implies solving a system  of  the
coupled differential equations, which,  unfortunately,  lies  beyond
our abilities.

To proceed further  we  will  not   calculate   the  amplitude,  but
consider instead  a  more  simple  problem. Namely, we want to  find
out under which conditions this amplitude will not be small.

We use the following terminology. Let the  energy,  $E$,  be  fixed.
Consider a path (\ref{35}) that lies  entirely  in  the  classically
allowed region:
\be
U[K(r),\lambda]<E,\ \  \lambda\in [0,\pi].               \label{48}
\ee
Such a path will be called {\it an  overbarrier  path}.  If  a  path
passes also through the classically forbidden region then  we  shall
call it an {\it underbarrier path}, for the evolution  along  such a
path  implies barrier penetration.

For an underbarrier paths the amplitude (\ref{46}) is small,  as  it
includes the small tunneling factor. This allows us to exclude  from
the sum (\ref{47}) the contribution of all  underbarrier  paths,  as
this contribution  is  certainly small. The remaining  sum  over the
overbarrier paths will not be small if only this sum includes a term
(or terms),  ${\cal  A}_{K(r)}$,  whose  value  is  stationary  with
respect to small variations  of  $K(r)$. Such a term corresponds  to
the contribution of a stationary  phase  path. Thus,  the  amplitude
(\ref{47}) will not be small provided  there  exists  a   stationary
phase path on the set  of  all  overbarrier  paths.  Therefore,   to
proceed further, we first need  to  select  the  overbarrier   paths
from the set (\ref{35}), and next to look  for  a  stationary  phase
path among them.

To select the overbarrier paths in accordance with Eq.(\ref{48}) one
has to know the properties of the potential $U[K(r),\lambda]$.
It  is  useful  to  treat  the  functional   $U[K(r),\lambda]$    in
geometrical terms, viewing it as an infinite dimensional surface  in
the corresponding function space. Let us  call  this   surface  {\it
an energy  surface}  or {\it a potential  barrier  surface}.  It  is
natural to call the directions  in  this  surface    generated    by
$\partial/\partial\lambda$    and   $\partial/ \partial K(r)$   {\it
a  transverse  direction}  and  {\it  a   longitudinal   direction}
respectively.  Indeed,  changing  of  $\lambda$  with  fixed  $K(r)$
corresponds to motion across  the  barrier  towards  a  neighbouring
vacuum. Changing of $K(r)$ implies passing  to  other   paths,  i.e.
the motion along  the   barrier.   When   $K(r)$   is   fixed,   the
potential    reaches     a     maximum      at      $\lambda=\pi/2$,
so  we  will  call the functional
\be \varepsilon[K(r)]=U[K(r),\lambda=\pi/2]                \label{49}
\ee
{\it a barrier height  functional}  and  say  that  it  defines  the
profile of the top of the barrier.

Consider the critical points of the energy surface. It is clear that
such points have to  belong  to  the  top,  $\lambda=\pi/2$.  Direct
variation of  the  functional  (\ref{49}) yields \cite{main}
\be
\delta\varepsilon=2\iii\{-((1-\frac{2m}{r})\sigma
K')'+\frac{K(K^{2}-1)} {r^{2}}\sigma\}\delta K  dr,      \label{50}
\ee
where $\delta K(0)=\delta  K(\infty)=0$,  functions  $m,\sigma$  are
given by Eq.(\ref{33}) ($\lambda=\pi/2$). The  vanishing   of   this
variation implies the condition
\be
((1-\frac{2m}{r})\sigma K')'=\sigma\frac{K(K^{2}-1)}{r^{2}}.
                                                        \label{51}
\ee
Next note that when  $\lambda=\pi/2$,  functions  $m$  and  $\sigma$
given by Eq.(\ref{33}) obey the following equations
\be
\sigma'=2\frac{K'^{2}}{r}\sigma,\ \ \ \ \ \ \ \ \ \
 (m\sigma)'=(K'^{2}+\frac{(K^{2}-1)^{2}}{2r^{2}})\sigma. \label{B5}
\ee
It  is  worth  noting  that  Eqs.(\ref{51}),(\ref{B5})  admit   such
solutions when the EYM fields (\ref{22}), (\ref{13})  coincide  with
the solutions of the EYM equations found  by  Bartnik  and  McKinnon
\cite{9}. These BK solutions are labeled by  an  integer,  $n$.  The
function $K(r)$ for the $n$-th solution has  $n$  zeros  and  it  is
usually denoted by $w_{n}$. For odd values of $n$ $w_{n}$ obeys  the
conditions  (\ref{23}).  The   masses   of   these   solutions   are
$M_{n}=\sqrt{4\pi}M_{pl}m_{n}/g$, where $M_{pl}$ being Planck's mass
and  $m_{n}$  increases  as  $n$  grows  from  the   minimal   value
$m_{1}=0.828$ to $m_{\infty}=1$. Thus  the  (odd-$n$)  BK  solutions
relate  to  critical  points  of  the  potential   barrier   surface
separating distinct vacua in the EYM theory. This  circumstance  has
allowed  us  to  interpret  in  Ref.\cite{10}  these  solutions   as
sphalerons (on  more  general  grounds  the  correspondence  between
extrema  of  the  energy  and   the   action   was   considered   in
Ref.\cite{21}).

It is clear now that if the energy  $E$  exceeds  the  mass  of  the
$n$-th EYM sphaleron, then the path passing  through  this  sphaleron
(such a path is defined  by  Eqs.(\ref{35})-(\ref{33})  with  $K(r)=
w_{n}(r)$) as well as neighbouring paths will be overbarrier.

Next we are looking for the stationary phase  path  on  the  set  of
these overbarrier paths. Our aim is to show that  the  path  passing
through the sphaleron will be such a  stationary  phase  path.  This
means that the quantum phase $\Phi[K(r)]$ defined  by  Eq.(\ref{46})
with $K(r)=w_{n}(r)$ should be stationary on variations  of  $K(r)$.
To check the stationarity of the phase let us consider an  arbitrary
variation of $K=w_{n}$:
\be
K(r)=w_{n}(r)+\alpha\f(r),                        \label{var}
\ee
where $\alpha$ is the variational parameter,  and  the  perturbation
obeys  $\f(r)=O(r^{2})$  as   $r\rightarrow0$,   $\f(r)=O(1/r)$   as
$r\rightarrow\infty$.    Inserting     this   into     $\Phi[K(r)]$,
$U[K(r),\lambda]$     and     $\varepsilon[K(r)]$     defined     by
Eqs.(\ref{45}),(\ref{46}),  (\ref{49}),  one   obtains   the   phase
$\Phi_{\varphi}(\alpha)$, (index $\varphi$ refers to the  choice  of
the perturbation), and the one and two-dimensional sections  of  the
energy  surface:  $\varepsilon_{\varphi}   (\alpha)$,   $U_{\varphi}
(\alpha,\lambda)$. The phase  $\Phi[K(r)]$  will  be  stationary  if
$\Phi_{\varphi}(\alpha)$ has an extremum for {\it  any}  $\f(r)$  in
Eq.(\ref{var}).

One should note that, strictly speaking, the exact stationary  phase
path is not contained in the path family (\ref{35}). Finding such  a
path implies solving the  general  equations  of  motion  \cite{11},
which  lies  beyond  the  scope  of  our  present  analysis.   Paths
(\ref{35})  specified  by  $K=w_{n}$  may  be   only   approximately
stationary. This means that for  these  approximate  paths  and  any
independent perturbations $\varphi(r)$ in (\ref{var}), the positions
of   the   extremum   of   the   corresponding    phase    functions
$\Phi_{\varphi}(\alpha)$ do not coincide exactly with  the  position
of the sphaleron, $\alpha=0$, however they are  close  together.  In
other  words,  functional  derivatives  $\left.  \delta   \Phi[K(r)]
\right|_{K=w_{n}}$ although do not vanish exactly, are  nevertheless
small.

Consider first such perturbations $\f(r)$ which increase the barrier
height:  $\varepsilon(\alpha\neq   0)>\varepsilon(0)$;   there   are
infinitely many such perturbations. One may see that  the  phase  is
stationary with respect to all these perturbations. Indeed, in  this
case  the  corresponding  two-dimensional  section  of  the   energy
surface, $U_{\varphi}(\alpha,\lambda)$, has the typical saddle shape
shown in Fig.1.  The  saddle  negative  $\lambda$-direction  on  the
picture  (shown  by  the  horizontal  arrow)  specifies   {\it   the
transverse rolling down mode} of  the  sphaleron  \cite{22}.  It  is
clear from this picture that the path passing through the  sphaleron
is the minimal potential energy path interpolating between  distinct
vacua, so it corresponds to an extremum of the phase. The analogous
situation takes place in the electroweak theory where the sphaleron
is  the
highest-energy point on a minimal-energy  path  connecting  distinct
vacua, which implies the stationarity of the phase for such
a path \cite{5}. Direct numerical inspection shows that, for several
ground state sphaleron energy-increasing perturbation modes $\f (r)$
checked,  the  phase  $\Phi_{\f}(\alpha)$  indeed  has  an  extremum
(minimum) in the vicinity of zero value of $\alpha$.

However, in the EYM theory,
contrary to the situation in  the electroweak case,
not  all  perturbations  (\ref{var})  increase  the  barrier
height, for sphalerons possess also {\it longitudinal}  negative
modes. Indeed, direct calculation  \cite{main}  shows  that  in  the
vicinity of a sphaleron the  following expansion holds
\be
\varepsilon [w_{n}(r)+\varphi(r)]=
\varepsilon[w_{n}(r)]+\delta^{2}\varepsilon +\ldots ,
                                                       \label{52}
\ee
where the first order term vanishes and dots denote higher order
terms. The second order term reads
\be
\delta^{2}\varepsilon =  \iii\f(-\frac{d^{2}}{dr_{\ast}^{2}}  +V)\f\
dr_{\ast}, \label{53}
\ee
with the new radial coordinate, $r_{\ast}$,  being  defined  by  the
following   relations    $\frac{dr}{dr_{\ast}}=\sigma(1-2m/r)$,    $
r_{\ast}(r=0)=0$. The effective potential is
\be
V=\sigma^{2}(1-\frac{2m}{r})\{\frac{3w_{n}^{2}-1}{r^{2}} +
\frac{8}{r^{3}}w_{n}'w_{n}(w_{n}^{2}-1)-
\frac{4}{r^{2}}w_{n}'^{2}(1-\frac{(w_{n}^{2}-1)^{2}}{r^{2}})\},
                                                         \label{54}
\ee
where functions $m$ and $\sigma$ relate to  corresponding  sphaleron
solutions, the derivatives are calculated with respect to  $r$.  One
may see that the scalar product for different perturbation modes can
be naturally defined as $<\f_{1},\f_{2}>=\int\f_{1}\f_{2}dr_{\ast}$;
independent modes being  orthogonal.  It  is  obvious  that  if  the
differential operator in Eq.(\ref{53}) has negative eigenvalues,
\be
(-\frac{d^{2}}{dr_{\ast}^{2}}+V)\     \f\     =\omega^{2}\f,\      \
\omega^{2}<0,                                         \label{55}
  \ee
corresponding   eigenfunctions,  $\f$,  will  specify   the   energy
decreasing perturbation. It  is  worth   noting  that  Eq.(\ref{55})
coincides (up to a  constant multiplier)  with  that  first  derived
by Straumann and Zhou  \cite{23}  in  their  analysis  of the linear
stability of the BK solutions, provided  that  one  identifies   the
quantity $\omega$  with  the time frequency  of   perturbations   of
the background  BK solution. It is known \cite{24},  that  when  the
potential  $V$  in  Eq.(\ref{55}) relates to the $n$-th BK solution,
the  equation  has  exactly   $n$  independent  negative  eigenvalue
solutions. Thus, the  $n$-th  EYM  sphaleron  has,  apart  from  one
transverse negative modes, also  $n$  longitudinal  negative  modes.
Inserting such perturbation modes in Eq.(\ref{var}), one obtains the
corresponding two-dimensional sections of the energy  surface  which
have the typical shape shown in Fig.2. One may see from this picture
that, for such a section, the path passing through the sphaleron  is
not the minimal energy path. So,  with  respect  to  these  negative
perturbations, the stationarity of the phase is not obvious  (it  is
usually assumed that a sphaleron  has  to  have  one  and  only  one
negative  mode  in  order  to  be  significant  for  the  transition
processes \cite{yaffe}). Nevertheless, we are  able  to  demonstrate
the stationarity of the phase also in this case, at  least  for  the
ground state $(n=1)$ sphaleron.

For the ground  state  sphaleron  there  exists  only  one  negative
eigenmode to Eq.(\ref{55}). It  turns  out  that  the  corresponding
perturbation (\ref{var}) may be related  to  the  rescaling  of  the
sphaleron field. Namely, instead of Eq.(\ref{var}) let
\be  K(r)=w_{1}(\beta r),                             \label{56}
\ee
where  $\beta$  is  the  scaling  parameter.   Inserting   this   in
Eq.({\ref{45}) one obtains the following two dimensional section  of
the energy surface:
\be
U(\beta,\lambda)=
\beta\sin^{2}\lambda\iii (w_{1}'^{2}+ \sin^{2}\lambda
\frac{(w_{1}^{2}-1)^{2}}{2r^{2}})exp(-2\beta^{2}
sin^{2}\lambda\int_{r}^{\infty}w_{1}'^{2}\frac{dr}{r})dr, \label{57}
\ee
(here $w_{1}=w_{1}(r)$), the value $U(1,\pi/2)$ being the  sphaleron
mass. This function tends to zero when  $\beta\rightarrow  0,\infty$
as well  as  when  $\lambda\rightarrow  0,\pi$.  The  plot  of  this
function is depicted in Fig.2. This picture demonstrates  explicitly
the existence of the two independent sphaleron negative  modes.  One
such mode  is  the  transverse  rolling  down  mode  \cite{22}  (the
$\lambda$-direction on the picture), the other is  the  longitudinal
negative mode \cite{23}  (the  $\beta$-direction).  (Note  that  the
infinitesimal scaling mode,  $\left.  \f=\partial_{\beta}w_{1}(\beta
r)\right|_{\beta=1}=rw'_{1}(r)$,  is not the exact eigenmode for the
Eq.(\ref{55}), but rather  a  superposition  of  the  true  negative
eigenmode and some other mode).

The scaling behaviour  of  the  quantum  phase  is  defined  by  the
inserting (\ref{56}) into (\ref{46}):
\be
\Phi_{scale}(\beta)=\int_{0}^{\pi}
d\lambda \sqrt{2\mu(\beta,\lambda)[E-U(\beta,\lambda)]},
                                                     \label{phase}
\ee
where $U[\beta,\lambda]$ is given by Eq.(\ref{57}) and
\be
\mu(\beta,\lambda)=
\frac{1}{\beta}\iii \frac{r^{2}}{\sigma}
(w_{1}'^{2}+2sin^{2}\lambda\
\frac{(w_{1}^{2}-1)^{2}}{r^{2}-2\beta mr\ })dr,      \label{60}
\ee
with
$$
\sigma(r)=exp\{-2\beta^{2}sin^{2}\lambda\int_{r}^{\infty}w_{1}'^{2}
\frac{dr}{r}\}, $$
\be
m(r)=\frac{\beta\sin^{2}\lambda}{\sigma(r)}\int_{0}^{r}(w_{1}'^{2}+
sin^{2}\lambda\frac{(w_{1}^{2}-1)^{2}}{2r^{2}})\sigma dr.\label{59}
\ee
Now, one  may  see  that  the  phase  (\ref{phase})  indeed  has  an
extremum.  First  note  that  when  $\beta$  is  small,  the   phase
$\Phi_{scale}(\beta)\sim  1/\sqrt{\beta}$,  i.e.  it  diverges  when
$\beta\rightarrow 0$ (remember  that  the  energy  $E$  exceeds  the
sphaleron mass, that is, the maximal value of the potential drawn in
Fig.2). Also, and this is the crucial  point,  $\Phi_{scale}(\beta)$
diverges when $\beta$ tends to some value $\beta_{c}=1.465$, because
the integral entering Eq.(\ref{60})  diverges  in  this  limit.  The
value $\beta=\beta_{c}$ corresponds  to  such  a  rescaling  of  the
sphaleron field when the rescaled configuration begins to acquire an
event horizon (all of the paths (\ref{35}),(\ref{56})  with  $\beta>
\beta_{c}$ pass through virtual black holes).  At  the  horizon  the
quantity $r^{2}-2\beta mr$ entering the denominator in Eq.(\ref{60})
vanishes, so the integral diverges as does the phase. So,  somewhere
in between, $0<\beta<\beta_{c}$, the phase must have a minimum  (see
Fig.3).

Thus we can see that when the energy exceeds the  ground  state  EYM
sphaleron mass, the  path  passing  through  the  sphaleron  insures
extremality  of  the  quantum  phase  with  respect  to  any   small
variations inside the path family (\ref{35}). The existence of  this
extremum means that the sum in Eq.(\ref{47}),  being  evaluated  via
stationary phase approximation, will  include the contribution of a
stationary point. In other words, there is the constructive
interference on the set of the overbarrier paths (\ref{35}), which
has to lead to an enhancement of the transition.

It turns out that the analysis carried  out  above  for  the  ground
state sphaleron remains valid for  the  higher (odd $n>1$) sphalerons.
This means that each higher sphaleron introduces a stationary  phase
path, only if the phase will be stationary  also  with
respect  to  the  variations corresponding to the
additional  $n-1$  longitudinal  higher  sphaleron
negative modes, which is very plausible. Thus, if the energy exceeds
the  masses  of  all  the  EYM  sphalerons  (i.e.   $E>1$   in   the
dimensionless units used), the sum  in  Eq.(\ref{47})  includes  the
contributions of (infinitely) many stationary points (one point  for
each higher sphaleron), which may lead to an additional  enhancement
of the transition rate.

When the energy is less then the ground state  sphaleron  mass,  the
situation changes. The overbarrier paths in this case also exist, as
can be seen in Fig.2. However,  there  exists  no  stationary  phase
path,  so  the constructive interference in absent.

All this allows us to conclude that in the extreme high energy limit
the effects of gravity have to lead to an enhancement of the
intensity of the winding  number changing
transitions. This means that the  rate  of  the
{\it inclusive} \cite{7} fermion number violating reactions at  very
high energies is not small. Such reactions lead to the formation  of
intermediate sphaleron states, the decay of  which  will  produce  a
large number of gravitons and gauge  bosons  \cite{25}  as  well  as
extra fermions due to the anomaly. It has been  argued  recently  by
Gibbons and Steif \cite{26}, that the  decay  of  the  BK  particles
should be accompanied by  fermion number non-conservation.
Our  arguments  show
that, at high energies, the probability of BK particles  being  born
and subsequently decaying is not small.  Notice  that  the  energies
available are very large -- the masses of the EYM sphalerons are  of
the order of $M_{pl}/g$. Such processes may arise  naturally  within
the context  of  superstring  theory  leading  to  a  ``primordial''
fermion asymmetry. Indeed, lower order terms of the expansion of the
superstring action in the string tension, give rise to  the  coupled
Einstein-Yang-Mills-Dilaton  (EYMD)  equation   of   motion.   These
equations  possess  classical  solutions,  which  resemble  in  many
respects the BK solutions and coincide  exactly  with  them  in  the
vanishing dilatonic coupling limit \cite{27}.  It  is  very  likely,
that these EYMD particles may  also  be  interpreted  as  sphalerons
\cite{corn,28}, and our present analysis may  be  extended  to  that
case.

\section*{Acknowledgments}

I would like to thank Professor  Norbert  Straumann  for  a  careful
reading of the manuscript  and also Professor Petr Hajicek for a
helpful critique.

This work was supported by  the  Swiss  National  Science
Foundation.

\newpage

\vspace{3 cm}

\begin{center}
{\bf Figure captions}
\end{center}
\vspace{2 cm}

{\bf Fig.1} Typical  shape  of  the  energy-surface  two-dimensional
section $U_{\f}(\lambda,\alpha)$  for  energy  increasing  sphaleron
perturbation modes $\f$. The  surface  forms  a  barrier  separating
distant EYM  vacua  ($\lambda=0,\pi$).  Position  of  the  sphaleron
($\alpha=0,\lambda=\pi/2$) is shown by the vertical arrow.

{\bf Fig.2} Plot  of  the  function  $U(\beta,\lambda)$  defined  by
Eq.(\ref{57}).  Vertical  arrow   shows   the   sphaleron   position
$(\beta=1,\lambda=\pi/2)$.  Horizontal  arrows  correspond  to   the
transverse    ($\lambda$-direction)     and     the     longitudinal
($\beta$-direction) sphaleron negative modes.

{\bf Fig.3} Behaviour of  the  quantum  phase  $\Phi_{scale}(\beta)$
defined by Eq.(\ref{phase}) with $E=1$. The path passing through the
sphaleron is specified by the value $\beta=1$.

\end{document}